\documentclass[12pt]{article}

\usepackage{amsmath}
\usepackage{amssymb}
\usepackage{slashed}
\usepackage[numbers,sort&compress]{natbib}
\usepackage{hyperref}
\usepackage{cleveref}
\crefname{equation}{Eq.}{Eqs.}
\crefname{figure}{Fig.}{Figs.}
\crefname{table}{Table}{Tables}
\crefname{section}{Section}{Sections}

\def\non{\nonumber\\}

\usepackage[letterpaper,margin=1in,bottom=1in]{geometry}
\usepackage{float} 
\usepackage{parskip} 
\usepackage{tabulary} 
\usepackage{color} 
\usepackage{soul} 
\usepackage{subfigure}
\usepackage{graphicx}
\usepackage[section]{placeins} 

\def\lhc2{LHC~Run~II}

\usepackage{cleveref}

\def\.4{\vspace{-.5cm}}

\def\beq{\begin{equation}}
\def\be{\begin{equation}}
\def\beqn{\begin{eqnarray}}
\def\ee{\end{equation}}
\def\eeq{\end{equation}}
\def\eeqn{\end{eqnarray}}

\author{
Pran Nath\footnote{Email: p.nath@northeastern.edu}~\ and 
Maksim Piskunov\footnote{Email:m.piskunov@northeastern.edu}\\~\\
Department of Physics, Northeastern University,
Boston, MA 02115-5000, USA
}

\title{Supersymmetric Dirac-Born-Infeld Axionic Inflation and Non-Gaussianity}

\begin{document}
\maketitle
\date

\textbf{Abstract: } 
  An analysis is given of inflation based on a supersymmetric Dirac-Born-Infeld (DBI) action in
  an axionic landscape. The DBI model we discuss involves a landscape of chiral superfields with one
  $U(1)$ shift symmetry which is broken by instanton type non-perturbative terms
  in the superpotential. Breaking of the shift symmetry leads to one pseudo-Nambu-Goldstone-boson which acts as
  the inflaton while the remaining normalized phases of the chiral fields
  generically labeled axions are invariant under the
  $U(1)$ shift symmetry. The analysis is carried out in the vacuum with stabilized saxions,
 which are the magnitudes of
  the chiral fields. Regions of the parameter space where slow-roll inflation occurs are exhibited and the spectral
  indices as well as the ratio of the tensor to the scalar power spectrum are computed. An interesting aspect of supersymmetric
  DBI models analyzed is that in most of the parameter space tensor to scalar ratio and scalar spectral index are consistent with
   Planck data if     slow roll occurs and is not eternal. Also interesting is that
  the ratio of the tensor to the scalar power spectrum can be large and can lie close to the
  experimental upper limit and thus testable in improved
  experiment. Non-Gaussianity in this
  class of models is explored.
\newpage

\section{Introduction}
  As is well known many of the problems associated with Big Bang cosmology which include
  the flatness problem, the horizon problem, and the monopole problem are resolved by inflation~\cite{Guth:1980zm,Starobinsky:1980te,Linde:1981mu,Albrecht:1982wi,Sato,Linde:1983gd}.
  Quantum fluctuations at the time of horizon exit carry significant information regarding specifics
  of the inflationary model~\cite{Mukhanov+,Cheung:2007st}
  which can be extracted from cosmic microwave background (CMB) radiation anisotropy.
  The data from the Planck experiment~\cite{Adam:2015rua,Ade:2015lrj,Array:2015xqh} has helped
  constrain inflation models excluding some and narrowing down the parameter space of others.
  One such model is so called natural inflation based on a $U(1)$ shift symmetry
  which is described by a simple potential~\cite{Freese:1990rb,Adams:1992bn}
  $V(a) = \Lambda^4 \left(1+ \cos(\frac{a}{f})\right)$,
  where $a$ is the axion field and $f$ is the axion decay constant.
  In this case consistency with Planck data requires the axion decay constant to be significantly greater than the Planck mass $M_\text{P}$.
  However, an axion decay constant larger than the Planck mass is undesirable since a global symmetry is not preserved by quantum gravity
  unless it has a gauge origin. Additionally string theory prefers the axion decay constant to lie below $M_\text{P}$~\cite{Banks:2003sx,Svrcek:2006yi}.
  It turns out that the reduction of the decay constant poses a problem, and several suggestions exist regarding its resolution such as the
  so called alignment mechanism \cite{Kim:2004rp,Long:2014dta}.

  A procedure for resolving the axion decay constant problem was proposed in~\cite{Nath:2017ihp}.
  One element of this analysis relies on a
  decomposition of the potential into a fast-roll and a slow-roll parts where the slow roll is controlled only by the inflaton field while the remaining fields enter in fast roll and are not relevant for
  inflation~\cite{Nath:2017ihp} (for a review of inflation in supersymmetric theories see, e.g., \cite{Nath:2016qzm}).
  An analysis within this model shows that one can obtain
  spectral indices as well as the ratio of the tensor to the scalar power spectrum consistent with the Planck data~\cite{Adam:2015rua,Ade:2015lrj,Array:2015xqh}.
  Another quantity of interest in primordial perturbations is the so-called non-Gaussianity~\cite{Maldacena:2002vr,Seery:2005wm,Seery:2005gb,Chen:2005fe,Chen:2006nt,Lyth:2005fi}. It is known that models
  with canonical kinetic energy do not lead to non-Gaussianity and for non-Gaussianity one needs models
  with non-canonical kinetic energy. In this context the Dirac-Born-Infeld (DBI) models are of interest (see, e.g.,~\cite{Alishahiha:2004eh,Easson:2007dh,Huang:2007hh,Gordon:2000hv,Langlois:2008wt}) which is the object of study in this work.
  Our work is focused on using shift symmetry and axions for inflation. For a partial list of other works where
  shift symmetry of axions is utilized in inflation
  see~\cite{ArkaniHamed:2003mz,Kaplan:2003aj,Green:2009ds,Higaki:2014pja,Higaki:2014mwa,Kadota:2016jlw,Kobayashi:2016vcx}
  and in the string context see~\cite{Kachru:2003sx,BlancoPillado:2004ns,Cicoli:2016olq}.
  For reviews of axionic cosmology see~\cite{Pajer:2013fsa,Marsh:2015xka}.\\

  The outline of the rest of the paper is as follows: In section \ref{sec2} we
  give a summary of previous results on the decomposition of a landscape of axion fields which undergo shifts
  under a $U(1)$ global transformation into fast-roll and slow-roll parts.
  This is one of the central elements in the
  analysis of the inflationary models we discuss later.
  In section \ref{sec3} we give a description of the supersymmetric DBI Lagrangian
  in superspace for the case of two chiral fields $\Phi_1$ and $\Phi_2$ which are oppositely charged under a $U(1)$ global symmetry.
  We then display the bosonic part of the Lagrangian after integration over the Grassmann co-ordinates.
  Here it is shown that the Lagrangian depends on the dimensionless parameters $\alpha_1, \alpha_2, \alpha_3$; and $T$
  which has the dimension of the fourth power of mass. The general case including $\alpha_1, \alpha_2, \alpha_3$
  is too complicated to discuss analytically and thus here the analysis is given taking into account only the $\alpha_1$ terms.
  In section \ref{sec4} we discuss the pressure, density and the inflation equations for a generic DBI model. In section \ref{sec5} we
  discuss the slow-roll parameters, non-Gaussianity, and the speed of sound which enters in defining non-Gaussianity.
  Model simulations and 
  experimental test of the two field DBI model is discussed in section \ref{sec6}, and conclusions are given in section \ref{sec7}.
  Further, details of the analysis are given in sections \ref{appenA}, \ref{appenB}, and \ref{appenC}.
 
\section{Fast-roll and slow-roll decomposition \label{sec2}}
  Before discussing the supersymmetric DBI model we summarize first the slow-roll and fast-roll decomposition of the
  inflation potential which is one of the central components of the analysis of this paper for the DBI case. As noted above
  the slow-roll and
  fast-roll decomposition of the potential was introduced in~\cite{Nath:2017ihp}.
  This analysis utilizes a landscape of pairs of chiral fields which are charged under a $U(1)$ global symmetry. Thus suppose we have
  a set of chiral fields $\Phi_i$ ($i=1, \cdots, n$) where $\Phi_i$ carry the same charge under the shift symmetry and the fields
  $\tilde \Phi_i$ ($i=1, \cdots, n$) carry the opposite charge. We assume that under $U(1)$ transformations the fields transform
  as follows
  \begin{align}
    \Phi_i\to e^{i q \lambda} \Phi_i, ~~\tilde \Phi_i\to e^{-i q \lambda} \tilde \Phi_i, ~~i=1, \cdots, n\,.
  \end{align}
  The superfields ${\Phi}_{i}$ have an expansion,
  \begin{align}
    {\Phi}_{i} = {\phi}_{i} + \theta {\chi}_{i} + \theta \theta {F}_{i}\,,
  \end{align}
  where ${\phi}_{i}$ is a complex scalar field consisting of the saxion (the magnitude) and the axion (the normalized phase), ${\chi}_{i}$ is the axino, and ${F}_{i}$ is an auxiliary field.
  Similarly the superfields $\tilde {\Phi}_{i}$ have an expansion:
  $\tilde {\Phi}_{i} = \tilde {\phi}_{i} + \tilde \theta \tilde {\chi}_{i} + \tilde\theta \tilde \theta \tilde{F}_{i}$.
  We may parametrize $\phi_i$ and $\tilde \phi_i$ so that
  \begin{align}
    \phi_i = \frac{1}{\sqrt 2}(f_i + \rho_i) e^{ia_i/f_i}, ~~~\tilde\phi_i = \frac{1}{\sqrt 2}(\tilde f_i + \tilde \rho_i) e^{i\tilde a_i/\tilde f_i}\,,
  \end{align}
  where $f_i= <\phi_i> ,~\tilde f_i= <\tilde\phi_i>$ and $(\rho_i, a_i)$ and $(\tilde \rho_i, \tilde a_i)$
  are the fluctuations of the quantum fields around their vacuum expectation values $f_i, \tilde f_i$.
  The above constitute
  $2n$ number of axionic fields $a_1, \cdots, a_n$ and $\tilde a_1, \cdots, \tilde a_n$.
  Since there is only one $U(1)$ shift symmetry, we can pick a basis where
  only one linear combination of it is variant under the shift symmetry and all others are
  invariant. We label this new basis $a_-, a_+, b_1, b_2, \cdots, b_{n-1}, \tilde b_1, \tilde b_2, \cdots, \tilde b_{n-1}$
  where only $a_-$ is sensitive to the shift symmetry. Thus the object of central interest is the field $a_-$ which
  is the pseudo-Nambu-Goldstone-Boson (pNBG) and acts
  as the inflaton. It can be expressed in terms of the original set of axion fields as below
  \begin{align}
    {a}_{-} &= \frac{1}{f_e} \left( \sum_{i= 1}^{m} {f}_{i} {a}_{i} - \sum_{i = 1}^{m} {\tilde{f}}_{i} {\tilde{a}}_{i} \right)\,.
    \label{combinations}
  \end{align}
  \begin{align}
    f_e= \sqrt{\sum_{i = 1}^{m} {f}_{i}^{2} + \sum_{i = 1}^{m} {\tilde{f}}_{i}^{2} }\,.
    \label{fe}
  \end{align}
  The relation Eq.~(\ref{fe}) was derived in~\cite{Nath:2017ihp} (see also~\cite{Ernst:2018bib}).
  The result of Eq.~(\ref{fe}) gives $f_e=\sqrt N f$ for the case when $f_i=\tilde f_i=f$ and $N=2m$
  which is the N-flation result but derived here in a different context~\cite{Dimopoulos:2005ac}.
 
\section{Supersymmetric DBI action for two chiral fields \label{sec3}}
  Supersymmetric DBI actions have been investigated by a number of authors
  (see, e.g., \cite{Khoury:2010gb,Khoury:2011da,Baumann:2011nk,Baumann:2011nm,Rocek:1997hi,Tseytlin:1999dj,Ito:2007hy,Billo:2008sp,Sasaki:2012ka,Aoki:2016tod}.
  Here we discuss the supersymmetric DBI in the context of axion inflation.
  The case of a single field DBI is given in section \ref{appenA}.
  Here we consider a pair of chiral superfields $\Phi_1$ and $\Phi_2$ which carry opposite
  charges under a global $U(1)$ symmetry.
  The supersymmetric Lagrangian involving $\Phi_1$ and $\Phi_2$ is given by
  \begin{equation}
    \mathcal{L}= \mathcal{L}_D+\mathcal{L}_{F},
    \label{1.1}
  \end{equation}
  where $\mathcal{L}_D$ is the D-part of the Lagrangian and $\mathcal{L}_{F}$ is the F-part. We consider the D-part consisting of
  a part $\mathcal{L}^{(1)}_D$ which is quadratic in the fields and a part $\mathcal{L}^{(2)}_D$ which is quartic in the fields so that
  \begin{align}
    \mathcal{L}_D = \mathcal{L}^{(1)}_D + \mathcal{L}^{(2)}_D,
    \label{Lag.9}
  \end{align}
  where $\mathcal{L}^{(1)}_D$ and $\mathcal{L}^{(2)}_D$ are invariant under the $U(1)$ symmetry and are given by
  \begin{align}
    \mathcal{L}^{(1)}_D
    &= \int d^4\theta \left(\Phi_1 \Phi_1^\dagger + \Phi_2 \Phi_2^\dagger \right)
    \label{Lag.10}
  \end{align}
  and
  \begin{align}
    \mathcal{L}^{(2)}_D=
    \mathcal{L}^{(2a)}_D+
    \mathcal{L}^{(2b)}_D+
    \mathcal{L}^{(2c)}_D+
    \mathcal{L}^{(2d)}_D+
    \mathcal{L}^{(2e)}_D,
  \end{align}
  where 
  \begin{align}
    \mathcal{L}^{(2a)}_D&=
      \int d^4\theta
      \frac{\alpha_1}{16T}\left(D^\alpha \Phi_1 D_\alpha \Phi_1\right)\left({\bar{D}}^{\dot{\alpha}}\Phi_1^\dagger {\bar{D}}_{\dot{\alpha}}\Phi_1^\dagger \right)
      G(\phi),\non
    \mathcal{L}^{(2b)}_D &=
      \int d^4\theta
      \frac{\alpha_1}{16T}\left(D^\alpha \Phi_2 D_\alpha \Phi_2\right)\left({\bar{D}}^{\dot{\alpha}}\Phi_2^\dagger {\bar{D}}_{\dot{\alpha}}\Phi_2^\dagger \right)
      G(\phi),\non
    \mathcal{L}^{(2c)}_D &=
      \int d^4\theta
      \frac{\alpha_2}{16T}\left(D^\alpha \Phi_1 D_\alpha \Phi_1\right)\left({\bar{D}}^{\dot{\alpha}}\Phi_2^\dagger {\bar{D}}_{\dot{\alpha}}\Phi_2^\dagger \right)
      G(\phi),\non
    \mathcal{L}^{(2d)}_D &=
      \int d^4\theta
      \frac{\alpha_2}{16T}\left(D^\alpha \Phi_2 D_\alpha \Phi_2\right)\left({\bar{D}}^{\dot{\alpha}}\Phi_1^\dagger {\bar{D}}_{\dot{\alpha}}\Phi_1^\dagger \right)
      G(\phi),\non
    \mathcal{L}^{(2d)}_D &=
      \int d^4\theta
      \frac{\alpha_3}{16T}\left(D^\alpha \Phi_1 D_\alpha \Phi_2\right)\left({\bar{D}}^{\dot{\alpha}}\Phi_1^\dagger {\bar{D}}_{\dot{\alpha}}\Phi_2^\dagger \right)
      G(\phi).
    \label{Lag.11}
  \end{align}
  Here
  \begin{align}
    G(\phi) = \frac{1}{T}\frac{1}{1+ A +\sqrt{(1+A)^2 -B}},
  \end{align}
  and $A$ and $B$ are assumed to have the following forms
  \begin{align}
    A &= (\partial_a\phi_1 \partial^a \phi^*_1 + \partial_a\phi_2 \partial^a \phi^*_2)/T, \non
    B &= \Big(
      \alpha_1\partial_a \phi_1 \partial^a \phi_1 \partial_b \phi^*_1 \partial^b \phi^*_1
      + \alpha_1 \partial_a \phi_2 \partial^a \phi_2 \partial_b \phi^*_2 \partial^b \phi^*_2
      + \alpha_2 \partial_a \phi_1 \partial^a \phi_1 \partial_b \phi^*_2 \partial^b \phi^*_2 \non
    & + \alpha_2 \partial_a \phi_2 \partial^a \phi_2 \partial_b \phi^*_1 \partial^b \phi^*_1
      + \alpha_3 \partial_a \phi_1 \partial^a \phi_2 \partial_b \phi^*_1 \partial^b \phi^*_2
    \Big) / T^2.
    \label{Lag.22}
  \end{align}
  We note that the Lagrangian of Eq.~(\ref{Lag.11}) is a direct generalization of the Lagrangian for the single field
  case (see section \ref{appenA}) which can be derived from a more basic 3-brane action 
  (see, e.g., \cite{Rocek:1997hi,Tseytlin:1999dj,Sasaki:2012ka} and the references
  therein). Here we simply extend the analysis to two fields
  in the most general supersymmetric form involving four covariant derivatives. In writing Eq.~(\ref{Lag.11}) we imposed
  an additional constraint which is invariance under $\Phi_1$ and $\Phi_2$ interchange.
  The possible relation of this Lagrangian
  to an underlying string model is an open question. Here we simply treat Eq.~(\ref{Lag.11}) as an effective low energy
  theory. Finally $\mathcal{L}_{F}$ is given by
  \begin{equation}
    \mathcal{L}_{F}=\int d^2\theta W\left(\Phi_1,\Phi_2\right)+\int d^2\bar\theta
    W^* \left(\Phi_1^\dagger,\Phi_2^\dagger\right),
  \end{equation}
  where the superpotential $W$ is given by
  \begin{equation}
    W=W_s+W_{sb}.
    \label{w1}
  \end{equation}
  Here $W_s$ is invariant under the global $U(1)$ symmetry and is taken to be of the form
  \begin{equation}
    W_s=\mu \Phi_1\Phi_2+\frac{\lambda}{2}{\left(\Phi_1\Phi_2\right)}^2\,.
    \label{w2}
  \end{equation}
  The form Eq.~(\ref{w2}) is chosen so that we can stabilize the saxion VEVs. Eq.~(\ref{w2}) also explains
  why we need a pair of chiral fields with opposite $U(1)$ charges because with a single chiral field which is
  charged under a $U(1)$ symmetry we cannot form a non-trivial $W_s$, needed for stabilizing the saxions,
  which is invariant under the
  $U(1)$ symmetry. $W_{sb}$ breaks the global $U(1)$ symmetry and is taken to be of the form
  \begin{equation}
    W_{sb}=\sum_{k=1}^m\left(A_{1, k}\Phi_1^k+A_{2, k}\Phi_2^k\right).
    \label{w3}
  \end{equation}
  We note in passing that the superpotential of the type Eqs. (\ref{w1})-(\ref{w3}) was considered in \cite{Nath:2017ihp,Halverson:2017deq}.
  Integration over $\theta's$ gives for the full Lagrangian
  \begin{align}
    \mathcal{L}
      &=   \mathcal{L}_D +  \mathcal{L}_F= \mathcal{L}_I + \mathcal{L}_{II},\non
    \mathcal{L}_I &= T - T \sqrt{(1+A)^2 - B},\non
    \mathcal{L}_{II}& = F_1 F^*_1 + F_2 F^*_2
      + G(\phi) \Big[ \alpha_1(-2 F_1 F^*_1 \partial_a\phi_1 \partial^a \phi^*_1
      + F_1^2 {F^*_1}^2)
      + \alpha_1 (-2 F_2 F^*_2 \partial_a\phi_2 \partial^a \phi^*_2 \non
    & + F_2^2 {F^*_2}^2)
      + \alpha_2 (-2 F_1 F^*_2 \partial_a\phi_1 \partial^a \phi^*_2
      + F_1^2 {F^*_2}^2 -2 F_2 F^*_1 \partial_a\phi_2 \partial^a \phi^*_1
      + F_2^2 {F^*_1}^2) \non
    & + \alpha_3(
      -F_1 F^*_1 \partial_a \phi_2 \partial^a \phi^*_2 - F_2 F^*_2 \partial_a \phi_1 \partial^a \phi^*_1
      + F_1 F_2 F^*_1 F^*_2) \Big]
      + \left(\frac{\partial W}{\partial \phi_1}F_1+ \frac{\partial W}{\partial \phi_2}F_2 + h.c.\right).
    \label{Lag.23}
  \end{align}
  There are four auxiliary fields in Eq.~(\ref{Lag.23}) which are $F_1, F^*_1, F_2, F^*_2$. The field equations obtained by varying
  $F_1, F^*_1, F_2, F^*_2$ are given in section \ref{appenB}. These are coupled equations
  involving all the $F$'s and $F^*$'s
  and solving them is non-trivial. Much simplicity results if we set $ \alpha_2=0=\alpha_3$.
  In this case as
  shown in section \ref{appenB} the auxiliary fields $F_k$ satisfy the cubic equation
  \begin{align}
    F_k^3+ p_k F_k + q_k=0\,, k=1,2\,,
  \end{align}
  where $p_k, q_k$ are defined by
  \begin{equation}\label{DisplayFormulaNumbered:eq.twoDBI.p.1}
  \begin{split}
    & p_k={\left(\frac{\partial W}{\partial \phi_k}\right)}^{-1}\frac{\partial W^*}{\partial \phi^*_k}\frac{1-2\alpha_1 G\left(\phi\right)\partial_\mu \phi_k\partial^\mu \phi_k}{2\alpha_1 G\left(\phi\right)}, \\
    & q_k=\frac{1}{2\alpha_1 G\left(\phi\right)}{\left(\frac{\partial W}{\partial \phi_k}\right)}^{-1}{\left(\frac{\partial W^*}{\partial \phi^*_k}\right)}^2. \\
  \end{split}
  \end{equation}
  Since $F_k$ satisfies a cubic equation, there are three roots which are given by
  \begin{equation}
  \begin{split}
    F_k=& \omega^j {\left(-\frac{q_k}{2}+\sqrt{ {\left(\frac{q_k}{2}\right)}^2+{\left(\frac{p_k}{3}\right)}^3}\right)}^{1/3}\\
    &+ \omega^{3-j}{\left(-\frac{q_k}{2}-\sqrt{ {\left(\frac{q_k}{2}\right)}^2+{\left(\frac{p_k}{3}\right)}^3}\right)}^{1/3},
  \end{split}
  \label{auxsolution}
  \end{equation}
  where $\omega$ is the cube root of unity and $j=0,1,2$.
Naively, it appears there are three solutions for $F_k$. However, as exhibited in section \ref{appenB}, only $j = 0$ is a solution to the full Euler-Lagrange equations for $F$.
  Setting the derivative terms to zero, the scalar potential of the theory for this case can be computed
  and is exhibited in section \ref{appenC}. As an expansion in $1/T$, the $F_k$ takes the form
  \begin{equation}
  \begin{split}
    & F_k=-\frac{\partial W^*}{\partial \phi^*_k}+\frac{1}{T} \left(\frac{\partial W}{\partial \phi_k}\right){\left(\frac{\partial W^*}{\partial \phi^*_k}\right)}^2-\frac{3}{T^2} {\left(\frac{\partial W}{\partial \phi_k}\right)}^2{\left(\frac{\partial W^*}{\partial \phi^*_k}\right)}^3+\frac{12}{T^3}{\left(\frac{\partial W}{\partial \phi_k}\right)}^3{\left(\frac{\partial W^*}{\partial \phi^*_k}\right)}^4 \\
    & \indent{}-\frac{55}{T^4}{\left(\frac{\partial W}{\partial \phi_k}\right)}^4{\left(\frac{\partial W^*}{\partial \phi^*_k}\right)}^5+\frac{273}{T^5}{\left(\frac{\partial W}{\partial \phi_k}\right)}^5{\left(\frac{\partial W^*}{\partial \phi^*_k}\right)}^6+\mathcal{O}\left(\frac{1}{T^6}\right).
  \end{split}
  \label{fexpand}
  \end{equation}
  Further the scalar potential when expanded in powers of $1/T$ takes the form
  \begin{equation}
  \begin{split}
    & V\left(\phi \right)=\sum_{k=1}^2\Big[\frac{\partial W}{\partial \phi_k}\frac{\partial W^*}{\partial \phi^*_k}-\frac{1}{2T}{\left(\frac{\partial W}{\partial \phi_k}\frac{\partial W^*}{\partial \phi^*_k}\right)}^2+\frac{1}{T^2}{\left(\frac{\partial W}{\partial \phi_k}\frac{\partial W^*}{\partial \phi^*_k}\right)}^3 \\
    & \indent{}-\frac{3}{T^3}{\left(\frac{\partial W}{\partial \phi_k}\frac{\partial W^*}{\partial \phi^*_k}\right)}^4+\frac{11}{T^4}{\left(\frac{\partial W}{\partial \phi_k}\frac{\partial W^*}{\partial \phi^*_k}\right)}^5-\frac{91}{2T^5}{\left(\frac{\partial W}{\partial \phi_k}\frac{\partial W^*}{\partial \phi^*_k}\right)}^6+\mathcal{O}\left(\frac{1}{T^6}\right)\Big].
  \end{split}
  \label{vexpand}
  \end{equation}
  Thus as $T \rightarrow \infty$ we recover the conventional results for $F_k$ and $V\left(\phi\right)$. Further, the above analysis also implies stability conditions so that
  \begin{align}
    \frac{\partial W}{\partial \phi_k} = 0= \frac{\partial W^*}{\partial \phi^*_k}, ~~k=1,2.
    \label{min-eq}
  \end{align}
  We use Eq.~(\ref{min-eq}) for stabilizing the saxions. Thus we parametrize $\phi_k$ so that
  \begin{equation} \label{DisplayFormulaNumbered:eq.twoDBI.phi}
    \phi_k = \frac{1}{\sqrt 2} \left(f_k+\rho_k\right) e^{i a_k/f_k}, k=1,2\,,
  \end{equation}
  where $\rho_k$ are the saxion fields, $a_k$ are the axions and $f_k$ are the axion decay constants.
  For the case of the assumed superpotential the
  stabilization conditions on a CP conserving vacuum are
  \begin{equation} \label{DisplayFormulaNumbered:eq.twoDBI.stabilization.1}
    \mu f_2 + \frac{1}{2} \lambda f_1 f_2^2+\sum_{k=1}^m \frac{1}{2^{k/2-1}} k A_{1, k}f_1^{k-1}=0,
  \end{equation}
  \begin{equation} \label{DisplayFormulaNumbered:eq.twoDBI.stabilization.2}
    \mu f_1 + \frac{1}{2} \lambda f_1^2 f_2+\sum_{k=1}^m \frac{1}{2^{k/2-1}} k A_{2, k}f_2^{k-1}=0.
  \end{equation}
  For the case of two fields, we have $F_1$ and $F_2$, which can be solved using Eq.~(\ref{auxsolution}).
  For our choice of $W$, we can evaluate $p_k$ and $q_k$ (k=1,2) explicitly so that we have
  \begin{equation} \label{DisplayFormulaNumbered:eq.twoDBI.p.1}
  \begin{split}
    & p_1
    =\frac{\mu \phi^*_2+\lambda \phi^*_1{\phi^*_2}^2+\sum_{k=1}^mk A_{1, k}{\phi^*_1}^{k-1}}{\mu \phi_2+\lambda \phi_1\phi_2^2+\sum_{k=1}^mk A_{1, k}\phi_1^{k-1}}\frac{1-2\alpha_1G\left(\phi\right)\partial_\mu \phi_1\partial^\mu \phi^*_1}{2\alpha_1G\left(\phi\right)},
  \end{split}
  \end{equation}
  \begin{equation} \label{DisplayFormulaNumbered:eq.twoDBI.p.2}
  \begin{split}
    & p_2
    =\frac{\mu \phi^*_1+\lambda {\phi^*_1}^2 \phi^*_2 + \sum_{k=1}^mk A_{2, k}{\phi^*_2}^{k-1}}{\mu \phi_1+\lambda \phi_1^2\phi_2+\sum_{k=1}^mk A_{2, k}\phi_2^{k-1}}\frac{1-2\alpha_1G\left(\phi\right)\partial_\mu \phi_2\partial^\mu \phi^*_2}{2\alpha_1G\left(\phi\right)},
  \end{split}
  \end{equation}
  \begin{equation} \label{DisplayFormulaNumbered:eq.twoDBI.q.1}
  \begin{split}
    & q_1
    =\frac{1}{2\alpha_1G\left(\phi_1\right)}\frac{{\left(\mu \phi^*_2 + \lambda \phi^*_1 {\phi^*_2}^2+\sum_{k=1}^mk A_{1, k}{\phi^*_1}^{k-1}\right)}^2}{\mu \phi_2+\lambda \phi_1\phi_2^2 + \sum_{k=1}^mk A_{1, k}\phi_1^{k-1}},
  \end{split}
  \end{equation}
  \begin{equation} \label{DisplayFormulaNumbered:eq.twoDBI.q.2}
  \begin{split}
    & q_2
    =\frac{1}{2\alpha_1G\left(\phi_2\right)}\frac{{\left(\mu \phi^*_1 + \lambda {\phi^*_1}^2 \phi^*_2 + \sum_{k=1}^mk A_{2, k}{\phi^*_2}^{k-1}\right)}^2}{\mu \phi_1+\lambda \phi_1^2\phi_2 + \sum_{k=1}^mk A_{2, k}\phi_2^{k-1}}.
  \end{split}
  \end{equation}
  In these equations we didn't impose stability conditions which must be imposed for evaluation.
  We now carry out a fast-roll and a slow-roll decomposition of the fields following the procedure discussed in \cite{Nath:2017ihp} and reviewed in \ref{sec2} and define
  \begin{align} \label{eq.twoDBI.aPlusMinusFull}
    a_+/f_+={\frac{1}{\sqrt 2}}\left(a_1/f_1+a_2/f_2\right),\non
    a_-/f_-={\frac{1}{\sqrt 2}}\left(a_1/f_1-a_2/f_2\right).
  \end{align}
  Here $a_+$ is the field that  is invariant under the shift symmetry and  undergoes fast roll and $a_-$ is the field that is sensitive to the shift symmetry and undergoes slow roll. We are interested in only the slow-roll part
  and thus we suppress $a_+$ and retain only the $a_-$ part in the potential. \\
  To simplify the analysis we  set $A_{1,k} = A_{2,k} = A_k$ and set $f_1 = f_2 = f_+ = f_- = f$.
  In this case we have neglected spatial gradients:
  \begin{align}
    A( a_-)& = -\frac{{\dot{a}_-}^2}{2T},
    ~~~B(a_-) = \frac{1}{8 T^2}  {\dot a_-}^4,
    \label{const3}
  \end{align}
  \begin{equation}
    G_i\left(a_-\right)= G\left(a_-\right)
    = \frac{1}{ T- \dot a_-^2/2 + \sqrt{T^2- T \dot a_-^2 + \frac{1}{8} {\dot a}_{-}^4 \left(2 - \alpha_1\right)}}\,.
    \label{const5}
  \end{equation}
  $\mathcal{L}_I$ is given by
  \begin{align} \label{eq.axionLagrangian.I}
    \mathcal{L}_{I} &= T \left(1-\sqrt{1-\frac{{\dot a_{-}}^2}{T}+\frac{\left(2-\alpha_1\right){\dot a_{-}}^4}{8 T^2}} \right).
  \end{align}
  $\mathcal{L}_{II}$ is more complicated:
\begin{equation}
  \begin{split} \label{eq.axionLagrangian.II}
    \mathcal{L}_{II} &= T \left(2\mathcal{F}_{+}^2+2\mathcal{F}_{-}^2-\frac{4}{3\alpha_1}\left(\mathcal{T}+\left(\alpha_1-1\right)\frac{{\dot{a}}_{-}^2}{4 T}\right) + 4k \left(\mathcal{F}_{+}+\mathcal{F}_{-}\right) \right. \\
    & \left.{}+\frac{\alpha_1}{\mathcal{T}-{\dot{a}}_{-}^2/\left(4T\right)}\left(2\left(\mathcal{F}_{+}^2+\mathcal{F}_{-}^2-\frac{2}{3\alpha_1}\left(\mathcal{T}+\left(\alpha_1-1\right)\frac{{\dot{a}}_{-}^2}{4 T}\right)\right)\frac{{\dot{a}}_{-}^2}{4 T} +\mathcal{F}_{+}^4+\mathcal{F}_{-}^4 \right.\right. \\
    & \left.\left.{}+\frac{2}{3\alpha_1^2}{\left(\mathcal{T}+\left(\alpha_1-1\right)\frac{{\dot{a}}_{-}^2}{4 T}\right)}^2-\frac{4}{3\alpha_1}\left(\mathcal{T}+\left(\alpha_1-1\right)\frac{{\dot{a}}_{-}^2}{4 T}\right)\left(\mathcal{F}_{+}^2+\mathcal{F}_{-}^2\right)\right)\right),
  \end{split}
  \end{equation}
  where
  \begin{equation}
    \mathcal{T}=\frac{1}{2} \left(1+\sqrt{1-\frac{{\dot{a}}_{-}^2}{T}+\frac{\left(2-\alpha_1\right){\dot{a}}_{-}^4}{8 T^2}}\right),
  \end{equation}
  \begin{equation}
    k=\tilde\beta \sqrt{\sum_{m, n}m n {\cal{G}}_m {\cal {G}}_n \left(1-\cos\left(\frac{a_- m}{\sqrt 2 f}\right)-\cos\left(\frac{a_- n}{\sqrt 2 f}\right)+\cos\left(\frac{a_- \left(m-n\right)}{\sqrt 2 f}\right)\right)},
  \end{equation}
  \begin{equation}
    \mathcal{F}_{\pm} = \pm \left(\mp\frac{1}{2\alpha_1}k \left(\mathcal{T}-\frac{{\dot{a}}_{-}^2}{4 T}\right) + \sqrt{ \frac{1}{4\alpha_1^2} k^2 \left(\mathcal{T}-\frac{{\dot{a}}_{-}^2}{4 T}\right)^2+\frac{1}{27\alpha_1^3}{\left(\mathcal{T}+\left(\alpha_1-1\right)\frac{{\dot{a}}_{-}^2}{4 T}\right)}^3}\right)^{1/3},
  \end{equation}
  \begin{equation}
    {\cal{G}}_k = \frac{A_k 2^{1/2 (1 - k)}}{\tilde \beta \sqrt T f^{1-k}}\,.
  \end{equation}
  Here $\tilde \beta$ is an arbitrary dimensionless parameter which we choose such that ${\cal{G}}_k \sim 1$, and which determines the scale of symmetry breaking terms relative to $T$.
  

\section{Pressure, density and inflation equations \label{sec4}}
  The pressure $p$ and density $\rho$ are defined in terms of the stress tensor by
  \begin{equation}\label{DisplayFormulaNumbered:eq.pressure}
    p= \frac{1}{3} \sum_{i=1}^3 T^{ii}\,,
    ~~\rho =T^{00},
  \end{equation}
  where $T^{\mu\nu}$ $(\mu=0, 1, 2, 3)$ is the stress tensor and is given by
  \begin{equation}
    T^{\mu \nu}=g^{\mu \nu} \mathcal{L}\left(\phi_k,\partial_\mu \phi_k \partial^\mu \phi_k \right)-2\frac{\delta \mathcal{L}\left(\phi_k,\partial_\mu \phi_k \partial^\mu \phi_k \right)}{\delta \left(\partial_\mu \phi_k \partial^\mu \phi_k \right)}\partial^\mu \phi_k \partial^\nu \phi_k,
    \label{tmunu}
  \end{equation}
  and we use the metric $\eta^{\mu\nu}= {\rm diag}(-1, 1, 1, 1)$.
  In the analysis we assume space to be homogeneous and isotropic, so that $\partial_i\phi_k = 0$ for $1 \leq i \leq 3$ and 
	$\partial_0 \phi_{k} = \dot{\phi}_k$,
  one finds that Eqs. (\ref{DisplayFormulaNumbered:eq.pressure}) and (\ref{tmunu})
  for pressure and density become:
  \begin{equation}\label{DisplayFormulaNumbered:eq.pressure.cosmology}
    p = \mathcal{L}\left(\phi_k, - \beta_k \right),
  \end{equation}
  and
  \begin{equation}\label{DisplayFormulaNumbered:eq.density.cosmology}
    \rho = -\mathcal{L}\left(\phi_k, - \beta_k\right) + 2\sum_{k}\frac{\delta \mathcal{L}\left(\phi_k, - \beta_k \right)}{\delta \beta_k}
    \beta_k,
  \end{equation}
  where $\beta_k = \dot \phi_k^2$.
  These relations are valid for non-canonical kinetic terms.
  The Friedman equations are given by
  \begin{align}
    \dot \rho = -3H (p + \rho),
    \label{fried1}
  \end{align}
  \begin{align}
    3 M_\text{P}^2 \frac{\dot R^2}{R^2} = \rho,
    \label{fried2}
  \end{align}
  \begin{align}
    \dot\rho= - 6H \sum_{k} \frac{\partial L}{\partial \beta_k} \beta_k\,.
    \label{rhodot1}
  \end{align}
  Further taking the time derivative of Eq.~(\ref{DisplayFormulaNumbered:eq.density.cosmology})
  which along with Eq.~(\ref{rhodot1}) gives
  \begin{align}
    &\sum_{k}\Big[ \left(2 \frac{\partial^2 L}{\partial \beta_k \partial \beta_k} \beta_k + \frac{\partial L}{\partial \beta_k}
    \right)\dot \beta_k
    - \frac{\partial L}{\partial \phi_k} \dot \phi_k + 6H \frac{\partial L}{\partial \beta_k} \beta_k
    + 2 \frac{\partial^2 L}{\partial \phi_k\partial \beta_k} \beta\dot \phi_k \Big]=0.
    \label{motion}
  \end{align}
  Next we focus on the slow-roll part where we keep only the field $a_-$. In this case Eqs. (\ref{fried2}) and (\ref{motion}) can be written in
  the following form
  \begin{align}
    3 M_\text{P}^2 \frac{\dot R^2}{R^2} = 2 \dot a_{-}^2 \frac{\partial L}{\partial \dot a_{-}^2} - L\,,
    \label{inf-1a}
  \end{align}
  \begin{align}
    2 \left[2 \frac{\partial^2 L}{\partial \dot a_{-}^2 \partial \dot a_{-}^2} \dot a_{-}^2 + \frac{\partial L}{\partial \dot a_{-}^2}
    \right] \ddot a_{-} -
    \frac{\partial L}{\partial a_{-}} + 6H \dot a_{-} \frac{\partial L}{\partial \dot a_{-}^2}
    + 2 \frac{\partial^2 L}{\partial a_{-} \partial \dot a_{-}^2} \dot a_{-}^2
    =0
    \label{inf-2a}\,.
  \end{align}
  For the case of canonical kinetic energy and no dependence of the potential on time derivative of field, i.e.,
  \begin{align}
    L= -\frac{1}{2} \partial_\mu a_{-}\partial a_{-}^{\mu} - V(a_-)
  \end{align}
  Eqs. (\ref{inf-1a}) and  (\ref{inf-2a}) 
   reduce to the following
  \begin{align}
    3 M_\text{P}^2 \frac{\dot R^2}{R^2} = \frac{1}{2} \dot a_{-}^2 + V(a_-)\,,\non
    \ddot a_{-} + 3 H \dot a_{-} + V'(a_{-})=0\,,
  \end{align}
  which are correctly the relation for slow roll for the case when one has canonical kinetic energy and the potential
  is velocity independent. However, in our case we have more terms.

\section{Slow roll parameters and non-Gaussianity \label{sec5}}
  For non-canonical kinetic energy terms and for velocity dependent potential a quantity that enters
  the analysis of slow roll parameters is the speed of sound $c_s$ defined by
  \begin{align}
    c_s^2= \frac{p,\beta}{\rho,\beta}\,.
  \end{align}
  $c_s$ also enters in the analysis of non-Gaussianity to be discussed later.
  The speed of sound is limited by the constraint $0< c_s^2 \leq 1$. Often a parameter $\gamma$ is used which is defined by
  $\gamma = \frac{1}{c_s}$ and lies in the range $1 \leq \gamma < \infty$. For models with canonical kinetic energy $\gamma=1$
  and in this case there is no non-Gaussianity. For non-Gaussianity one requires $\gamma >1$. For the models we consider
  one may write $\gamma^2$ as follows
  \begin{align}
    \gamma^2
    &= 1+ 2 \dot a_{-}^2 \frac{\partial^2 L}{\partial \dot a_{-}^2 \partial \dot a_{-}^2}/\frac{\partial L}{\partial \dot a_{-}^2}
  \end{align}
  and Eq.~(\ref{inf-2a}) can then be written as follows
  \begin{align}
    \frac{\partial L}{\partial( \dot a_{-}^2/2)} \ddot a_{-}
    - \frac{1}{\gamma^2}
    \frac{\partial L}{\partial a_{-}} + \frac{3}{\gamma^2} H \dot a_{-} \frac{\partial L}{\partial (\dot a_{-}^2/2)}
    + \frac{1}{\gamma^2} \frac{\partial^2 L}{\partial a_{-} \partial (\dot a_{-}^2/2)} \dot a_{-}^2
    =0\,.
    \label{inf-2d}
  \end{align}
  The first three terms on the left hand side are similar to what one has normally except for $\gamma$ dependence.
  The last term on the left hand side is new.


  We define the slow roll parameters for DBI as~\cite{Maldacena:2002vr,Seery:2005wm,Seery:2005gb,Chen:2005fe,Chen:2006nt,Lyth:2005fi}
  \begin{align}
    \epsilon= -\frac{\dot H}{H^2}\,,~~\eta= \frac{\dot \epsilon}{\epsilon H}\,,~~
    s= \frac{\dot c_s}{c_s H}\,.
  \end{align}
  In terms of these parameters the power spectrum for the scalar perturbations $P_k^{\zeta}$
  and the power spectrum for the tensor perturbations $P_k^h$ are given by~\cite{Garriga:1999vw,ArmendarizPicon:1999rj}:
  \begin{align}
    P_k^{\zeta}&= \frac{1}{8\pi^2 M_\text{P}^2} \frac{H^2}{c_s \epsilon}\,,\non
    P_k^h &= \frac{2}{3\pi^2} \frac{\rho}{M_\text{P}^4}\,.
  \end{align}
  Further, the spectral indices $n_s$ and $n_t$ in this case are given by
  \begin{align}
    n_s&= 1-2 \epsilon -\eta -s\,,\non
    n_t&=- 2\epsilon\,,
    \label{SL1}
  \end{align}
  and the ratio $r$ of the tensor to the scalar power spectrum is \cite{Garriga:1999vw}
  \begin{align}
    r=\frac{P_k^h}{P_k^{\zeta}} &= - 8 c_s n_t\,.
    \label{SL2}
  \end{align}
  One may compare it to the conventional slow-roll parameters $\epsilon_V, \eta_V$
  for the case of the canonical kinetic energy term and no velocity dependence in the potential.
  Here one defines $\epsilon_V, \eta_V$ so that
  \begin{align}
    \epsilon_V= \frac{M_\text{P}^2}{2} \left(\frac{V'}{V}\right)^2\,,\non
    \eta_V= M_\text{P}^2 \frac{V^{''}}{V}\,,\non
  \end{align}
  and the spectral indices and the ratio of the tensor and the scalar power spectrum in this case are given by
  \begin{align}
    n_s=1- 6 \epsilon_V+ 2 \eta_V\,,~~~
    n_t= - 2 \epsilon_V\,,~~ r= 16 \epsilon_V\,.
    \label{SL3}
  \end{align}
  To establish a connection between the DBI slow-roll parameters $\epsilon, \eta$ with the conventional slow-roll parameters we note
  that in the conventional slow roll one assumes dominance of the potential and one sets $c_s=1$ and makes the following approximations
  \begin{align}
    \dot \phi\simeq - \frac{V'}{3H}\,,
    ~~H^2\simeq \frac{V}{3M_\text{P}^2}\,,~~H'= \frac{V'}{6M_\text{P}^2 H}\,.
    \label{approx1}
  \end{align}
  Using these one can connect $\epsilon, \eta$ to $\epsilon_V, \eta_V$ so that
  \begin{align}
    \epsilon&= \epsilon_V\,,\non
    \eta &= -2 \eta_V + 4 \epsilon_V\,.
    \label{SL4}
  \end{align}
  Using $c_s=1$ and Eq.~(\ref{SL4}) in Eqs. (\ref{SL1}) and (\ref{SL2}) we can recover Eq.~(\ref{SL3}).


  Non-Gaussianity is defined by the three-point correlation function of perturbations involving
  three scalars, two scalars and a graviton, two gravitons and a scalar and three gravitons~\cite{Maldacena:2002vr}.
  Since the work of~\cite{Maldacena:2002vr}  there has been
  a significant number of further analyses (see, e.g., \cite{Acquaviva:2002ud,Creminelli:2003iq,
  Silverstein:2003hf,Gruzinov:2004jx,Alishahiha:2004eh,Chen:2005fe,Creminelli:2005hu,
  Seery:2005wm,Babich:2004gb,Lyth:2005fi,Chen:2006nt, Huang:2007hh,Langlois:2008wt}).
  The dominant non-Gaussianity
  arises from the correlation function of three scalar perturbations. Thus for scalar perturbation $\zeta(\vec k)$ non-Gaussianity
  is defined by
  \begin{align}
    \left< \zeta(\vec k_1) \zeta(\vec k_2) \zeta(\vec k_3) \right> = (2\pi)^7 \delta^3(\vec k_1+ \vec k_2+ \vec k_3) \frac{\sum_i k_i^3}{\prod_i k_i^3}
    \left[ -\frac{3}{10} f_{NL} (P_k^{\zeta})^2 \right],
  \end{align}
  where $P_k^{\zeta}$ is the scalar power spectrum and $f_{NL}$ is a measure of non-Gaussianity. For the specific case of equilateral triangle
  when $k_1=k_2=k_3$, $f_{NL}$ is given by~\cite{Chen:2006nt}
  \begin{align}
    f_{NL}= \frac{35}{108}\left(\frac{1}{c_s^2}-1\right) - \frac{5}{81} \left[ \left(\frac{1}{c_s^2} -1- 2\frac{z_2}{z_1}\right) + (3-2 {c}_1)z \frac{z_2}{z_1}\right] + {\cal O}(\epsilon).
  \end{align}
  where
  \begin{align}
    z_1= \beta L_{,\beta} + 2 \beta^2 L_{\beta\beta}\,,~~
    z_2= \beta^2 L_{,\beta \beta} + \frac{2}{3} \beta^3 L_{,\beta\beta\beta}\,,~~
    z= \frac{\dot z_2}{z_2 H}\,.
  \end{align}

\section{Model simulation and experimental test \label{sec6}}

  \begin{figure}
    \centering
    \includegraphics[width=1.0\textwidth]{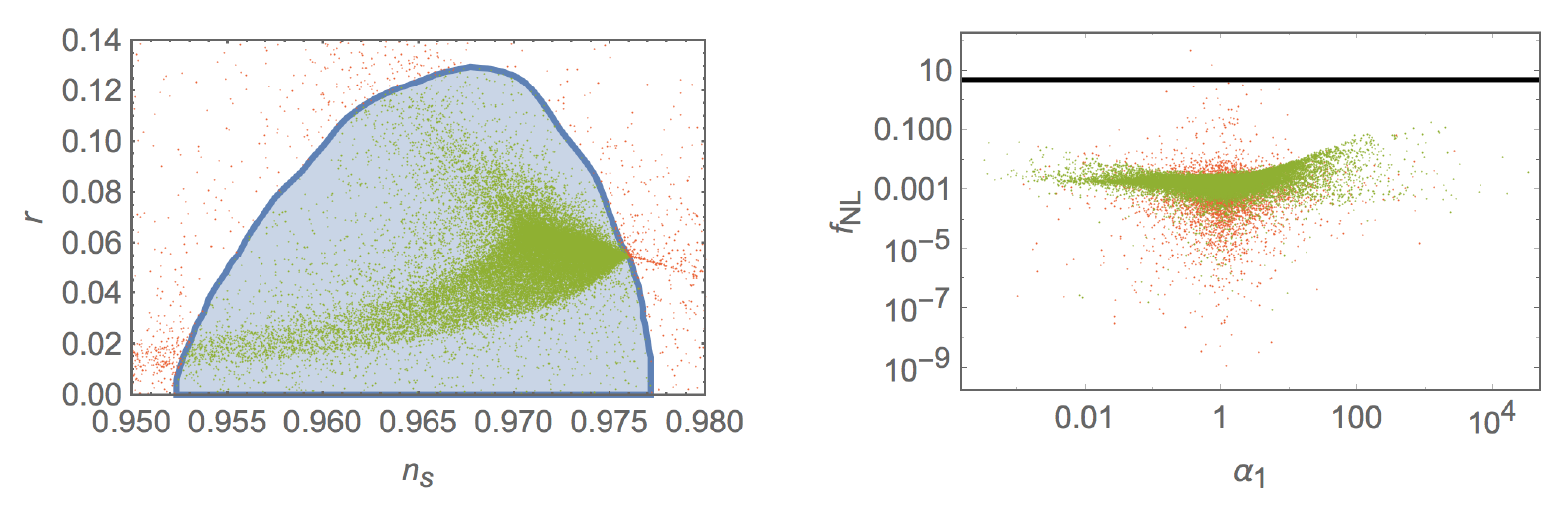}
    \caption{
      Left panel: A plot of the ratio $r$ of the tensor to scalar power spectrum vs the scalar spectral index $n_s$ 
      for the two field DBI model when $\alpha_1$ is allowed to vary, when $\alpha_2 = \alpha_3 = 0$. 
      The blue region enclosed by the blue line is the one allowed by experiment in the 2$\sigma$ range.      
            All the scatter points have pivot number of e-foldings in the range [50, 60].          
         $89\%$ of the simulated points 
      are consistent with Planck constraints as shown by green points while the red points lie outside the 68\% CL contour.
            Right panel: A plot of non-Gaussianity parameter $f_{NL}$ as
      a function of $\alpha_1$ for the same data set as in the left plot.  As in the left panel
      the green dots are parameter points which satisfy the Planck constraints
      on $r$ and $n_s$ and lie in the blue region  
      and the red dots are parameter points which are outside the experimentally allowed region  on $r$ and $n_s$. The horizontal black line is the
 lower limit $f_{NL} =5$ of the 
        projected value for observation in future experiments for non-Gaussianity.
           The points that lie above the black line give non-Gaussianity $f_\text{NL} > 5$ which, however, are not consistent with Eq.~(\ref{data}).
    }
    \label{fig1}
  \end{figure}

  \begin{figure}
    \centering
    \includegraphics[width=1.0\textwidth]{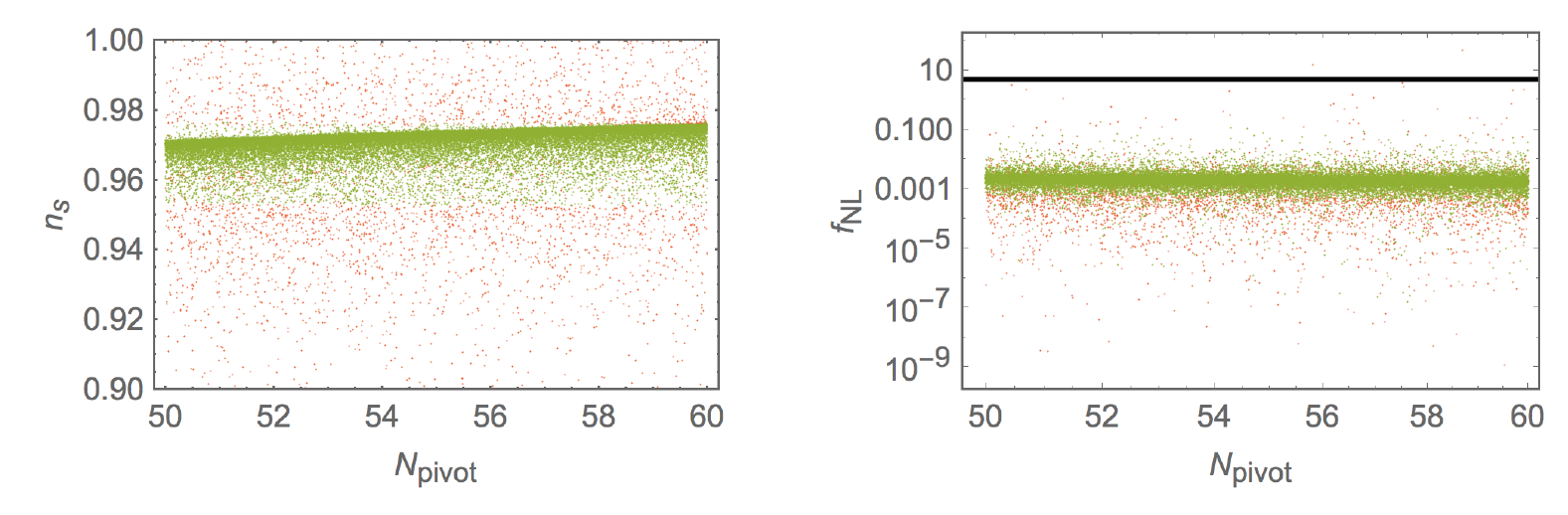}
    \caption{Left panel: A plot of the ratio $n_s$ vs $N_{\rm pivot}$ in the range $50-60$ for the same data set as in Fig.~(\ref{fig1}).
    The plot shows a mild dependence of $n_s$ on $N_{\rm pivot}$ in the range indicated. Right panel: The same as the left panel except
    $f_{NL}$ is plotted vs $N_{\rm pivot}$.
    The green and red dots are parameter points and have the same meaning as in Fig.~(\ref{fig1}).}
    \label{fig2}
  \end{figure}

  \begin{figure}
  	\centering
  	\includegraphics[width=1.0\textwidth]{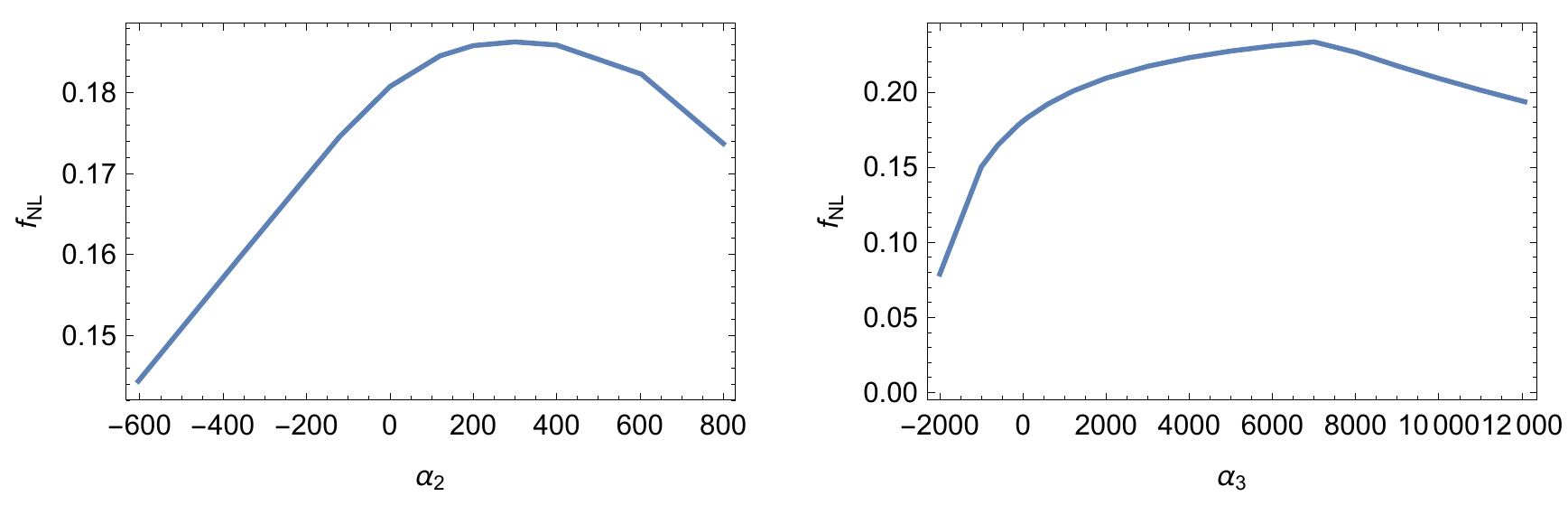}
  	\caption{The effect of $\alpha_2$ (left panel) and $\alpha_3$ (right panel) on non-Gaussianity for a parameter point 
	taken from Fig.~(\ref{fig1}) which is consistent with 
	Planck data and has the  largest non-Gaussianity among the set of parameter points 
  in Fig.~(\ref{fig1}). Specifically, here $f = 0.429 M_\text{P}$, $a_{-, 0} = 0.233 M_\text{P}$, $\dot a_{-, 0} = -0.118 M_\text{P}^2$, ${\cal{G}}_1 = {\cal{G}}_2 = {\cal{G}}_3 = 0$, ${\cal{G}}_4 = 1$, ${\cal{G}}_5 = -1.53$, ${\cal{G}}_6 = 563$, $\alpha_1 = 1207$, $\tilde \beta = 0.802$, $T = M_\text{P}^4$. One 
	finds that 
	 the effect is not sufficient to obtain a significantly larger value of $f_\text{NL}$.}
  	\label{fig3}
  \end{figure}

Numerical simulations performed consist of two parts: a large $0.5 \times 10^6$ points simulation in which $\alpha_2 = \alpha_3 = 0$, and a small simulation with $\alpha_2$ and $\alpha_3$ perturbed around an experimentally consistent point with the largest non-Gaussianity from the first simulation.
  The evolution of the inflaton field $a_-$ in the large simulation of Figs.~(\ref{fig1}) and (\ref{fig2}) is obtained by taking the effective axion Lagrangian Eqs.~(\ref{eq.axionLagrangian.I}) and (\ref{eq.axionLagrangian.II}), deriving Euler-Lagrange equation and Friedmann Eq.~(\ref{fried2}) using computer algebra, and solving them numerically. The integration is stopped when the ratio of the time derivative ($\dot n\left(t\right)$) of the e-foldings count to its time-average ($n\left(t\right) / t$) passes a threshold. Specifically,
  \begin{equation} 
  	\frac{\dot n \left(t\right)}{n\left(t\right) / t} < \theta_n,
  \end{equation} 
  where the threshold $\theta_n$ is chosen to be much less than one, and specifically
   in the analysis here it is chosen as $\theta_n = 1 / 16$. Increasing this parameter might stop integration prematurely, whereas decreasing it will worsen performance.
  An analytic expression similar to Eq.~(\ref{eq.axionLagrangian.II}), however,  cannot be derived  for  non-zero $\alpha_2$ and $\alpha_3$. In this case,  equations for the auxiliary fields $F_k$ are  solved numerically,  and the Lagrangian in terms of $\phi$ (Eq.~(\ref{Lag.23})) is
  also evaluated numerically for each set of values for ($a_-$, $\dot a_-$).  
   This process is significantly slower, therefore, in this case the simulations are only done for $29$ sets of parameter values as can be seen in Fig.~(\ref{fig3}). 
    The time of horizon exit is then found by counting $N_\text{pivot}$ e-foldings back from the end of integration, where $N_\text{pivot}$ is varied between $50$ and $60$. Note, the time of horizon exit depends on $\theta_n$, however, we find that due to $\dot n$ being small near the end of integration, the dependence is weak, and it affects the results less than the experimental uncertainty of $N_\text{pivot} \in \left[50, 60\right]$.{}

  Finally, we evaluate the experimental observables at the time of horizon exit using equations from section \ref{sec5} and reusing Euler-Lagrange and Friedmann equations to compute higher time derivatives of the Lagrangian, pressure and density. The observables computed are the ratio $r$ of the power spectrum of tensor to scalar perturbations, the spectral indices of scalar $n_s$ and tensor $n_t$ perturbations, 
 speed of sound $c_s$ 
  and the non-Gaussianity amplitude $f_\text{NL}$.
    In the $0.5 \times 10^6$ points Monte Carlo analysis we allow $\alpha_1$ to vary
  and search for solutions that satisfy the experimental constraints. Thus the
  current experimental limits from Planck experiment at $k_0=0.05\, {\rm Mpc}^{-1}$ are as follows~\cite{Adam:2015rua,Ade:2015lrj,Array:2015xqh}
  \begin{align}
  	n_s& = 0.9645 \pm 0.0049\, (68\% {\rm CL})\,, \non
  	r & <0.07\, (95\% {\rm CL})\,.
  	\label{data}
  \end{align}
  and the current experimental constraint on $c_s$ is~\cite{Ade:2015lrj}
  \begin{align}
  	c_s\geq 0.087~~~({\rm at ~}95\% ~CL)\,.
  	\label{sound}
  \end{align}
  The results of the analysis for the two field supersymmetric DBI discussed above
  are presented in Figs. (\ref{fig1}) and (\ref{fig2}). The left panel of Fig. (\ref{fig1}) gives
  a plot of $r$ vs $n_s$. Here the region enclosed by the blue line is the one allowed by experiment in the $2\sigma$ range.
  The Monte Carlo analysis shows that the experimentally allowed region is well populated by the parameter points of the model.
  One interesting feature of the analysis is that compared to the
  supersymmetric axion models discussed in~\cite{Nath:2017ihp}
  where $r$ was extremely small, here $r$ has significantly
  larger values. Thus the largeness of $r$ discriminates this class of supersymmetric DBI models from the models
  of \cite{Nath:2017ihp}.
  The right panel of Fig. (\ref{fig1}) gives a plot of $f_{NL}$ vs $\alpha_1$. We see a significant sensitivity
  of $f_{NL}$ to $\alpha_1$. However, overall $f_{NL}$ is typically small and mostly lies below 0.5. The effect of $\alpha_2$ and $\alpha_3$ for the point from Fig.~\ref{fig1} with largest non-Gaussianity is shown on Fig.~\ref{fig3}. One can see that although non-Gaussianity $f_\text{NL}$ is sensitive to $\alpha_2$ and $\alpha_3$, it still lies below 0.5. This level
  of non-Gaussianity appears too low for observation in the near future. The analysis of~\cite{Komatsu:2001rj,Komatsu:2003fd,Verde:1999ij}
  indicates that non-Gaussianity can be tested in data in future experiments provided $|f_{NL}|> 5$ and the $f_{NL}$
  given by the right panel of Fig. (\ref{fig1}) lies significantly below that. In the analysis we allow a corridor of $[50,60]$
  e-foldings and it is of interest to ask the dependence of $n_s$ and $f_{NL}$ on this number. In the left panel of Fig. (\ref{fig2})
  a plot of $n_s$ vs $N_{\rm pivot}$ is given. Here one finds that $n_s$ has a mild positive slope as a function of $N_{\rm pivot}$.
  The right panel gives a plot of $f_{NL}$ on $N_{\rm pivot}$. Here one finds the $f_{NL}$ has a relatively small variation in the range $[50,60]$ for
  $N_{\rm pivot}$. In either case one cannot draw any significant conclusion regarding the dependence of these parameters
  on the number of e-foldings as long as one is in the $N_{\rm pivot}$ range of $[50,60]$.

\section{Conclusion\label{sec7}}
	In this work we have analyzed inflation in a supersymmetric Dirac-Born-Infeld action with a $U(1)$ symmetry.
	Specifically we have carried out a detailed analysis of a pair of chiral DBI fields which possess opposite charges
	under the $U(1)$ symmetry. A transformation is then made to go to the co-ordinate frame where a linear
	combination of the axion fields is invariant under the global $U(1)$ transformations and an orthogonal combination
	is variant which acts as the inflaton. The $U(1)$ shift symmetry is broken by instanton type non-perturbative terms
	in the superpotential. The analysis is done
	in the vacuum state with stabilized saxions and the scalar potential can be decomposed into a fast-roll and a slow-roll
	part where the slow-roll part of the potential is now velocity dependent. This velocity dependence is a direct consequence of the
	DBI form of the action. In the analysis for the case $\alpha_2=0=\alpha_3$
	 we have obtained an explicit form for the auxiliary fields $F_k$ which satisfy a cubic equation
	in terms of the inflation field. We have analyzed the scalar and tensor power spectrum and computed the spectral indices.
	It is shown that a significant part of the parameter space exists which supports inflation consistent with the current experimental
	constraints on the ratio of the tensor to the scalar power spectrum and the spectral indices. A remarkable aspect of the
	proposed supersymmetric DBI model is that the model supports an observable value of the tensor to the scalar power spectrum
	and consequently also a significant value for the spectral index $n_t$. This is in contrast to a supersymmetric non-DBI model
	which typically has a suppressed value of $r$ and of $n_t$. An analysis of non-Gaussianity in the model was also carried out.
	It is found that for the model parameters that support inflation consistent with the Planck experimental values on $r$ and
	$n_s$, the non-Gaussianity is typically small. This holds true even when the parameters $\alpha_2$ and $\alpha_3$ along with
	$\alpha_1$ are included in the analysis.
		It is of interest to achieve this class of models in string theory using
	moduli stabilization of the type used in KKLT~\cite{Kachru:2003aw}
	or the Large Volume Scenario~\cite{Balasubramanian:2005zx}.

\textbf{Acknowledgments:}
	This research was supported in part by the NSF Grant PHY-1620575.

\section{Appendix A: Single field supersymmetric DBI Lagrangian \label{appenA}}

	To define notation and to discuss the technique used in the analysis of this work we consider the case here of a single
	superfield. Thus we consider a supersymmetric DBI Lagrangian of the form~\cite{Rocek:1997hi,Tseytlin:1999dj,Sasaki:2012ka}	
	\begin{equation}\label{DisplayFormulaNumbered:eq.dbiSuperLagrangian}
		\mathcal{L}_{DBI}=\int d^4\theta \left(\Phi \Phi^\dagger+\frac{1}{16}\left(D^\alpha \Phi D_\alpha \Phi \right)\left({\bar{D}}^{\dot{\alpha}}\Phi^\dagger {\bar{D}}_{\dot{\alpha}}\Phi^\dagger \right)G\left(\Phi \right)\right),
	\end{equation}
	where $\Phi$ and $\Phi^\dagger$ are the chiral and anti-chiral superfields, $D_\alpha$ and ${\bar{D}}_{\dot{\alpha}}$ are the supercovariant derivatives, and $G(\Phi)$ is given by
	\begin{equation}\label{DisplayFormulaNumbered:eq.dbiG} 
		G\left(\Phi \right)=\frac{1}{T}\frac{1}{1+A\left(\Phi \right)+\sqrt{{\left(1+A\left(\Phi \right)\right)}^2-B\left(\Phi \right)}}.
	\end{equation}
	Here $T$ is a parameter of the dimension of (mass)$^4$ and is related to the warp factor from the point of view of reduction of a ten dimensional theory to 4 dimensions. $A$ and $B$ are given by
	\begin{equation}\label{DisplayFormulaNumbered:eq.dbiA}
		A\left(\Phi \right)=\frac{\partial_\mu \Phi \partial^\mu \Phi^\dagger}{T},
		~~~B\left(\Phi \right)=\frac{\partial_\mu \Phi \partial^\mu \Phi \partial_\nu \Phi^\dagger \partial^\nu \Phi^\dagger}{T^2}.
	\end{equation}
	Ignoring fermions, the superfields have the expansion:
	\begin{equation}
	\begin{split}
		& \Phi \left(x,\theta,\bar{\theta}\right)=\Phi_L\left(x,\theta,\bar{\theta}\right) \\
		& =\phi \left(x\right)+\theta^\alpha \theta_\alpha F\left(x\right)+i \theta^\alpha {\sigma^\mu}_{\alpha \dot{\alpha}}{\bar{\theta}}^{\dot{\alpha}}\partial_\mu \phi \left(x\right)+\frac{1}{4}\theta^\alpha \theta_\alpha {\bar{\theta}}_{\dot{\alpha}}{\bar{\theta}}^{\dot{\alpha}}\partial^\mu \partial_\mu \phi \left(x\right),
	\end{split}
	\end{equation}
	\begin{equation}
	\begin{split}
		& \Phi^\dagger \left(x,\theta,\bar{\theta}\right)=\Phi_R\left(x,\theta,\bar{\theta}\right) \\
		& =\phi^*\left(x\right)+{\bar{\theta}}_{\dot{\alpha}}{\bar{\theta}}^{\dot{\alpha}}F^*\left(x\right)-i \theta^\alpha {\sigma^\mu}_{\alpha \dot{\alpha}}{\bar{\theta}}^{\dot{\alpha}}\partial_\mu \phi^*\left(x\right)+\frac{1}{4}\theta^\alpha \theta_\alpha {\bar{\theta}}_{\dot{\alpha}}{\bar{\theta}}^{\dot{\alpha}}\partial^\mu \partial_\mu \phi^*\left(x\right).
	\end{split}
	\end{equation}
	$\Phi \Phi^\dagger$ can be written in the form
	\begin{equation}\label{DisplayFormulaNumbered:eq.PhiPhiDg}
	\begin{split}
		& \Phi \Phi^\dagger =\phi \phi^*+\theta^2 \phi^* F+{\bar{\theta}}^2\phi F^* + i \theta \sigma^\mu \bar{\theta} \left(\phi^*\partial_\mu \phi -\phi \partial_\mu \phi^*\right) \\
		& \indent{}+\theta^2{\bar{\theta}}^2 \left(F F^*+\frac{1}{4}\phi \partial^\mu \partial_\mu \phi^*+\frac{1}{4}\phi^*\partial^\mu \partial_\mu \phi -\frac{1}{2}\partial^\mu \phi^*\partial_\mu \phi \right).
	\end{split}
	\end{equation}
	Further, since $\int d^4\theta \Phi \Phi^\dagger$ is the additive term in the Lagrangian, we can integrate by parts and get
	\begin{equation}\label{DisplayFormulaNumbered:eq.PhiPhiDgIntegral}
		\int d^2\theta d^2\bar{\theta} \Phi \Phi^\dagger
		= F F^*-\partial^\mu \phi^*\partial_\mu \phi.
	\end{equation}
	Next we note that
	\begin{equation}
		D_\beta \Phi =\left(\partial_\beta+i {\sigma^\mu}_{\beta \dot{\beta}}{\bar{\theta}}^{\dot{\beta}}\partial_\mu \right) \Phi.
	\end{equation}
	This leads to
	\begin{equation}\label{DisplayFormulaNumbered:eq.DPhi}
		D_\beta \Phi =-2\theta_\beta F\left(x\right)+2i {\sigma^\kappa}_{\beta \dot{\gamma}}{\bar{\theta}}^{\dot{\gamma}}\partial_\kappa \phi \left(x\right)+i \theta^\gamma \theta_\gamma {\sigma^\mu}_{\beta \dot{\beta}}{\bar{\theta}}^{\dot{\beta}}\partial_\mu F\left(x\right).
	\end{equation}
	Using the above we can compute $D^\alpha \Phi D_\alpha \Phi$ and the computation gives
	\begin{equation}\label{DisplayFormulaNumbered:eq.DPhiDPhi}
		D^\alpha \Phi D_\alpha \Phi =4 F^2\theta^2-4\partial^\kappa \phi \partial_\kappa \phi {\bar{\theta}}^2-8i F\partial_\kappa \phi \theta \sigma^\kappa \bar{\theta}-4 \partial^\mu F\partial_\mu \phi \theta^2 {\bar{\theta}}^2.
	\end{equation}
	The conjugate of Eq.~(\ref{DisplayFormulaNumbered:eq.DPhiDPhi}) can be computed as follows:
	\begin{equation}
		{\bar{D}}^{\dot{\alpha}}\Phi^\dagger {\bar{D}}_{\dot{\alpha}}\Phi^\dagger =- {\bar{D}}_{\dot{\alpha}}\Phi^\dagger {\bar{D}}^{\dot{\alpha}}\Phi^\dagger =-{\left(D^\alpha \Phi D_\alpha \Phi \right)}^\dagger,
	\end{equation}
	which gives
	\begin{equation}\label{DisplayFormulaNumbered:eq.DBPhiDgDBPhiDg}
	\begin{split}
		& {\bar{D}}^{\dot{\alpha}}\Phi^\dagger {\bar{D}}_{\dot{\alpha}}\Phi^\dagger
		=4{F^*}^2{\bar{\theta}}^2-4\partial^\kappa \phi^*\partial_\kappa \phi^* \theta^2+8i F^* \partial_\kappa \phi^* \theta \sigma^\kappa \bar{\theta}-4 \partial^\mu F^* \partial_\mu \phi^* \theta^2 {\bar{\theta}}^2.
	\end{split}
	\end{equation}
	The product $\left(D^\alpha \Phi D_\alpha \Phi \right)\left({\bar{D}}^{\dot{\alpha}}\Phi^\dagger {\bar{D}}_{\dot{\alpha}}\Phi^\dagger \right)$ in terms of component fields
	is given by
	\begin{equation}\label{DisplayFormulaNumbered:eq.DPhiDPhiDBPhiDgDBPhiDg}
	\begin{split}
		& \left(D^\alpha \Phi D_\alpha \Phi \right)\left({\bar{D}}^{\dot{\alpha}}\Phi^\dagger {\bar{D}}_{\dot{\alpha}}\Phi^\dagger \right)
		=16\left( F^2{F^*}^2+\partial^\kappa \phi \partial_\kappa \phi \partial^\mu \phi^*\partial_\mu \phi^*-2F F^*\partial^\mu \phi \partial_\mu \phi^*\right)\theta^2{\bar{\theta}}^2.
	\end{split}
	\end{equation}
	Note that $\left(D^\alpha \Phi D_\alpha \Phi \right)\left({\bar{D}}^{\dot{\alpha}}\Phi^\dagger {\bar{D}}_{\dot{\alpha}}\Phi^\dagger \right)$ already contains the highest possible power of the Grassmann numbers, therefore, the factor $G\left(\Phi \right)$ multiplying it in the case of the Lagrangian Eq.~(\ref{DisplayFormulaNumbered:eq.dbiSuperLagrangian}) can simply be replaced with $G\left(\phi \right)$.
	Combining terms we get the following expression for the Lagrangian:
	\begin{equation}\label{DisplayFormulaNumbered:eq.dbiComponentLagrangianNoAlgebra}
		\mathcal{L}_{DBI}=F F^*+G\left(\phi \right)\left(-2F F^*\partial^\mu \phi \partial_\mu \phi^* + F^2{F^*}^2\right)-T A\left(\phi \right)+G\left(\phi \right) B\left(\phi \right)T^2.
	\end{equation}
	We can further simplify and write the DBI Lagrangian in the form
	\begin{equation}\label{DisplayFormulaNumbered:eq.dbiComponentLagrangianNoW}
	\begin{split}
		\mathcal{L}_{DBI}
		& =-T\sqrt{1+2T^{-1}\partial_\mu \phi \partial^\mu \phi^* +{T^{-2}\left(\partial_\mu \phi \partial^\mu \phi^*\right)}^2-T^{-2} \left(\partial_\mu \phi \partial^\mu \phi \right)\left(\partial_\nu \phi^* \partial^\nu \phi^* \right)} \\
		& \indent{}+T+F F^* +G\left(\phi \right)\left(-2F F^* \partial^\mu \phi \partial_\mu \phi^* + F^2{F^*}^2\right).
	\end{split}
	\end{equation}

\section{Appendix B: Equations for $F_k, F^*_k ~(k=1,2)$ \label{appenB}}

	We need to eliminate $F_k$ and $F^*_k$ from the Lagrangian of Eq.~(\ref{Lag.23})
	to construct an equation of motion for $\phi_k$ only. In order to do that, we first vary with respect to $F_k$ and $F^*_k$ to obtain the following equations for $F_k$ and $F^*_k$. Here variations with respect to $F^*_1$ and $F_1$ give
	\begin{align}
		F_1 {+ \frac{\partial W^*}{\partial \phi_1^*}} + G(\phi) &\Big[
			   \alpha_1( -2 F_1 \partial_a\phi_1 \partial^a \phi_1^* + 2 F_1^2 F^*_1)
		     + \alpha_2( -2 F_2 \partial_a\phi_2 \partial^a \phi_1^* + 2F_2^2 F^*_1) \non
			&{+ \alpha_3( - F_1 \partial_a\phi_2 \partial^a \phi_2^* + F_1F_2 F^*_2)
		\Big] = 0\,,}\non
		F^*_1 {+ \frac{\partial W}{\partial \phi_1}} + G(\phi) &\Big[
			   \alpha_1( -2 F^*_1 \partial_a\phi_1 \partial^a \phi_1^* + 2 F_1 {F^*_1}^2)
			 + \alpha_2( -2 F^*_2 \partial_a\phi^*_2 \partial^a \phi_1 + 2{F^*_2}^2 F_1) \non
			&{+ \alpha_3( - F^*_1  \partial_a\phi^*_2 \partial^a \phi_2 + F^*_1F_2 F^*_2)
		\Big] = 0\,.}
	\end{align}
	Similarly variations with respect to $F^*_2$ and $F_2$ give
	\begin{align}
		F_2 {+ \frac{\partial W^*}{\partial \phi_2^*}} + G(\phi) &\Big[
			   \alpha_1(- 2 F_2 \partial_a \phi_2 \partial^a \phi_2^* + 2 F_2^2 F^*_2)
			 + \alpha_2(- 2 F_1 \partial_a \phi_1 \partial^a \phi_2^* + 2 F_1^2 F^*_2) \non
			&{+ \alpha_3(-   F_2 \partial_a \phi_1 \partial^a \phi_1^* + F_2 F_1 F^*_1)
		\Big] = 0\,,} \non
		F^*_2 {+ \frac{\partial W}{\partial \phi_2}} + G(\phi) &\Big[
			   \alpha_1(- 2 F^*_2 \partial_a \phi_2 \partial^a \phi_2^* + 2 F_2 {F^*_2}^2)
			 + \alpha_2(- 2 F^*_1 \partial_a \phi^*_1 \partial^a \phi_2 + 2 {F^*_1}^2 F_2) \non
			&{+ \alpha_3(-   F^*_2 \partial_a \phi^*_1 \partial^a \phi_1 + F^*_2 F_1 F^*_1)
		\Big] = 0\,.}
	\end{align}
	These give rise to a set of four coupled equations for $F_1, F^*_1, F_2, F^*_2$ which are difficult to solve analytically.
	To keep the analysis under control we set $\alpha_2=0=\alpha_3$. In this case the equations for
	$F_1, F^*_1$ become decoupled from those for $F_2, F^*_2$ and we get
	\begin{equation}
		{F^*_k}+\frac{\partial W}{\partial \phi_k}+\alpha_1G\left(\phi \right)\left(-2{F^*_k}\partial^\mu \phi_k \partial_\mu {\phi^*_k}+2F_k{{F^*_k}}^2\right)=0, ~~k=1,2\,,
		\label{barFk}
	\end{equation}
	\begin{equation}
		F_k+\frac{\partial W^*}{\partial {\phi^*_k}}+\alpha_1G\left(\phi \right)\left(-2F_k\partial^\mu \phi_k \partial_\mu {\phi^*_k}+2F_k^2{F^*_k}\right)=0\,, ~~k=1,2\,.
		\label{Fk}
	\end{equation}
	We multiply Eq.~(\ref{barFk}) by $F_k$ and Eq.~(\ref{Fk}) by $F^*_k$, and subtract one from another from which we can extract an
	equation for $F^*_k$ in terms of $F_k$:
	\begin{equation}\label{DisplayF_kormulaNumbered:eq.dbiF_kBF_kromF_k} 
		{F^*_k}=\left(\frac{\partial W}{\partial \phi_k}/\frac{\partial W^*}{\partial \phi^*_k}\right)F_k.
	\end{equation}
	Substitution back in Eq.~(\ref{Fk}) gives an equation for $F_k$:
	\begin{equation}\label{DisplayF_kormulaNumbered:eq.dbiF_kEquation}
		2\alpha_1G\left(\phi \right)\frac{\partial W}{\partial \phi_k}F_k^3+\frac{\partial W^*}{\partial \phi^*_k}\left(1-2\alpha_1G\left(\phi \right)\partial^\mu \phi_k \partial_\mu \phi^*_k\right)F_k+{\left(\frac{\partial W^*}{\partial \phi^*_k}\right)}^2=0.
	\end{equation}
	To simplify this equation we define $p$ and $q$ so that
	\begin{equation}\label{DisplayF_kormulaNumbered:eq.dbip}
		p_k={\left(\frac{\partial W}{\partial \phi_k}\right)}^{-1}\frac{\partial W^*}{\partial \phi^*_k}\frac{1-2\alpha_1G \partial_\mu \phi_k \partial^\mu \phi^*_k}{2\alpha_1G},
	\end{equation}
	\begin{equation}\label{DisplayF_kormulaNumbered:eq.dbiq}
		q_k=\frac{1}{2\alpha_1G}{\left(\frac{\partial W}{\partial \phi_k}\right)}^{-1}{\left(\frac{\partial W^*}{\partial \phi^*_k}\right)}^2.
	\end{equation}
	Substitution of $\partial^\mu \phi_k \partial_\mu \phi^*_k$ in terms of $p_k$ using Eq.~(\ref{DisplayF_kormulaNumbered:eq.dbip})
	in Eq.~(\ref{DisplayF_kormulaNumbered:eq.dbiF_kEquation})
	and using Eq.~(\ref{DisplayF_kormulaNumbered:eq.dbiq}) to eliminate ${\left(\frac{\partial W^*}{\partial \phi^*_k}\right)}^2$ in terms of
	$q_k$ in  Eq.~(\ref{DisplayF_kormulaNumbered:eq.dbiF_kEquation})
	gives
	\begin{equation}\label{DisplayF_kormulaNumbered:eq.dbiF_kpqEquation}
		F_k^3+p_k F_k+q_k=0,
	\end{equation}
	where we cancelled a common factor of $2G\frac{\partial W}{\partial \phi_k}$. The equation above is cubic in $F_k$ and it might appear that there are three consistent solutions.
	To see if this is the case we substitute Eq.~(\ref{DisplayF_kormulaNumbered:eq.dbiF_kBF_kromF_k}) back into Eq.~(\ref{auxsolution}) to obtain a solution for $F_k^*$. Upon simplification, one gets
	\begin{equation}
	\begin{split}
		F_k^*=& \omega^j {\left(-\frac{q_k^*}{2}+\sqrt{ {\left(\frac{q_k^*}{2}\right)}^2+{\left(\frac{p_k^*}{3}\right)}^3}\right)}^{1/3}\\
		&+ \omega^{3-j}{\left(-\frac{q_k^*}{2}-\sqrt{ {\left(\frac{q_k^*}{2}\right)}^2+{\left(\frac{p_k^*}{3}\right)}^3}\right)}^{1/3}.
	\end{split}
	\label{auxstarsolution}
	\end{equation}
	Note that this expression is only a complex conjugate of Eq.~(\ref{auxsolution}) if $j=0$ ($\omega^j$ and $\omega^{3-j}$ will have to be interchanged for it to be a complex conjugate in the case of $j = 1$ and $j = 2$), therefore only $j=0$ corresponds to a solution for the auxiliary fields $F_k$.

\section{Appendix C: DBI scalar potential with derivative terms absent \label{appenC}}
	To discuss the limit of the DBI potential to the standard supersymmetric potential we need to drop the derivative terms on $\phi$
	in the potential. In this case the full form of the scalar potential looks very different from the usual supersymmetric potential. To keep the
	expressions as simple as possible we consider the case of just one scalar field although extension to more fields is straightforward.
	In this case we have
	\begin{align}
		G&=\frac{1}{2T}\,,\non
		p&	=T {\left(\frac{\partial W}{\partial \phi}\right)}^{-1}\frac{\partial W^*}{\partial \phi^*},\non
		q&=T {\left(\frac{\partial W}{\partial \phi}\right)}^{-1}{\left(\frac{\partial W^*}{\partial \phi^*}\right)}^2,
	\end{align}
	$F$ is given by
	\begin{equation}\label{DisplayFormulaNumbered:eq.dbiPotentialF}
	\begin{split}
		& F=T^{1/2} \\
		& \indent{}\times{\left(\frac{\partial W}{\partial \phi}\right)}^{-1/3}{\left(\frac{\partial W^*}{\partial \phi^*}\right)}^{1/2}\left({\left(-\frac{1}{T^{1/2}}\sqrt{\frac{1}{4}\frac{\partial W^*}{\partial \phi^*}}+\sqrt{\frac{1}{27}{\left(\frac{\partial W}{\partial \phi}\right)}^{-1}+\frac{1}{4T}\frac{\partial W^*}{\partial \phi^*}}\right)}^{1/3}\right. \\
		& \indent\indent\left.{}+{\left(-\frac{1}{T^{1/2}}\sqrt{\frac{1}{4}\frac{\partial W^*}{\partial \phi^*}}-\sqrt{\frac{1}{27}{\left(\frac{\partial W}{\partial \phi}\right)}^{-1}+\frac{1}{4T}\frac{\partial W^*}{\partial \phi^*}}\right)}^{1/3}\right),
	\end{split}
	\end{equation}
	and the scalar potential is
	\begin{equation}\label{DisplayFormulaNumbered:eq.dbiPotentialWithF}
		V\left(\phi \right)=-\left(F^* F+\frac{\partial W}{\partial \phi}F+\frac{\partial W^*}{\partial \phi}F^*+\frac{1}{2T}F^2{F^*}^2\right).
	\end{equation}
	Further, an explicit form of the potential can be gotten by using
	Eq.~(\ref{DisplayFormulaNumbered:eq.dbiPotentialF}) back into the potential Eq.~(\ref{DisplayFormulaNumbered:eq.dbiPotentialWithF}) which gives
  \begin{equation}\label{DisplayFormulaNumbered:eq.dbiPotential}
    V\left(\phi \right) = V_1\left(\phi \right) + V_2\left(\phi \right) + V_3\left(\phi \right) + V_4\left(\phi \right),
  \end{equation}
  where
	\begin{equation}\label{DisplayFormulaNumbered:eq.dbiPotential}
		V_1\left(\phi \right)
		=-T {\left(\frac{\partial W}{\partial \phi}\frac{\partial W^*}{\partial \phi^*}\right)}^{1/6}\left(Q_{+}+Q_{-}\right)\left(Q_{+}^*+Q_{-}^*\right),
  \end{equation}
  \begin{equation}
    V_2\left(\phi \right)
		=-T^{1/2}{\left(\frac{\partial W}{\partial \phi}\right)}^{2/3}{\left(\frac{\partial W^*}{\partial \phi^*}\right)}^{1/2}\left(Q_+ + Q_-\right),
  \end{equation}
  \begin{equation}
		V_3\left(\phi \right) = V_2^* \left(\phi \right),
  \end{equation}
  \begin{equation}
		V_4\left(\phi \right)
    =-\frac{1}{2}T {\left(\frac{\partial W}{\partial \phi}\frac{\partial W^*}{\partial \phi^*}\right)}^{1/3}\left(Q_+ + Q_-\right)^2 \left(Q_+^* + Q_-^*\right)^2,
	\end{equation}
  and
  \begin{equation}
    Q_\pm = \left(-\frac{1}{T^{1/2}}\sqrt{\frac{1}{4}\frac{\partial W^*}{\partial \phi^*}}\pm\sqrt{\frac{1}{27}{\left(\frac{\partial W}{\partial \phi}\right)}^{-1}+\frac{1}{4T}\frac{\partial W^*}{\partial \phi^*}}\right)^{1/3}.
  \end{equation}
	We can expand $F$ given by  Eq.(\ref{DisplayFormulaNumbered:eq.dbiPotentialF}) 	
in powers of $1/T$ and this expansion is exhibited in Eq.~(\ref{fexpand}). Similarly we can expand $V$ given by 
Eq.(\ref{DisplayFormulaNumbered:eq.dbiPotentialWithF})  in powers of $1/T$ and this
	expansion is given in Eq.~(\ref{vexpand}).
	One can see that the lowest terms in the expansion for both $F$ and $V$ give the standard result.


\end{document}